\documentclass[conference]{IEEEtran}
\ifCLASSINFOpdf
   \usepackage[pdftex]{graphicx}
\else
   \usepackage[dvips]{graphicx}
\fi
\usepackage{amsmath}
\usepackage{array}
\usepackage{xcolor}
\usepackage{amsfonts}

\begin{document}
\title{Low-cost low-power in-vehicle occupant detection with mm-wave FMCW radar}

\author{
	\IEEEauthorblockN{Mostafa Alizadeh}
	\IEEEauthorblockA{Eletrical and Computer Engineering\\
		University of Waterloo\\
	Canada, Ontario, \\
	Email: m5alizad@uwaterloo.ca}
	\and 
	\IEEEauthorblockN{Hajar Abedi}
	\IEEEauthorblockA{System Design Engineering\\
	University of Waterloo \\
	Canada, Ontario,\\
	Email: habedifi@uwaterloo.ca}
	\and
	\IEEEauthorblockN{George Shaker}
	\IEEEauthorblockA{Eletrical and Computer Engineering\\
	University of Waterloo\\
	Canada, Ontario,\\
	Email: gshaker@uwaterloo.ca}
%\and
%\IEEEauthorblockN{George Shaker\\ and Montgomery Scott}
%\IEEEauthorblockA{Starfleet Academy\\
%San Francisco, California 96678--2391\\
%Fax: (888) 555--1212}
}


%Telephone: (800) 555--1212, Fax: (888) 555--1212}


\maketitle

\begin{abstract}
In this paper, we use a low-cost low-power mm-wave frequency modulated continuous wave (FMCW) radar for the in-vehicle occupant detection. We propose an algorithm using Capon filter for the joint range-azimuth estimation. Then, the minimum necessary features are extracted to train machine learning classifiers to have reasonable computational complexity while achieving high accuracy. In addition, experiments were carried out in a \textit{minivan} to detect occupancy of each row using support vector machine (SVM). Finally, our proposed system achieved 97.8\% accuracy on average in finding the defined scenarios. Moreover, The system can correctly identify if the vehicle is occupied or not with 100\% accuracy. 

%the performance of SVM model, in term of accuracy, revealed that our proposed system achieves at least 93% accuracy for the localization. 
\end{abstract}


% \fi
\IEEEpeerreviewmaketitle

\section{Introduction}
Frequency modulated continuous wave (FMCW) radars have unique advantages that they differentiate them from other radar systems. The benefits include simultaneous detection of range, Doppler (or velocity), and angle which makes this type of radar popular for a variety of applications. The major advantages of these radars are being low-cost and low-power so that make them suitable for most internet-of-things (IoT) applications such as
%remote glucose monitoring \cite{Glucose, Glucose1},
vital signs detection \cite{alizadeh}, wireless finger print identification \cite{karly_1}, gesture recognition \cite{karly_2}, and occupant detection without imposing any potential long-term health risks.

The sensor imaging resolution is defined as the minimum spatial separation of two targets revolvable by the radar which contains both range and azimuth resolutions. For FMCW radars, the larger the sweeping bandwidth is, the more range resolution is \cite{ours_ieee_access}. On the other hand, the angle of arrival resolution increases by increasing the number of transmitters and receivers. However, the implementation of a radar system with a large number of transmitters and receivers would result in higher system cost and more operational complexity. Hence it is desirable to achieve accurate occupancy detection with low-resolution radars. For instance, authors in \cite{Principal_COMP} used a single transceiver ultra-wideband (UWB) radar with only 5.35 cm without angle of arrival estimation. Indeed, Santra \textit{et al.} in \cite{infineon} used a mm-wave FMCW radar with 50 cm resolution at 1 m distance to the radar. In \cite{Principal_COMP}, they applied principle component analysis (PCA) on the \textit{range profile} for counting people in a room and the first component was only considered. The range profile is represents peaks corresponding to the object distributions in the scene and its first basis may not contain all target locations reducing the system performance. In \cite{infineon}, they tried to find human activities by categorizing them in three classes of macro-Doppler, micro-Doppler, and vital signs without invoking the azimuth information. In contrast in our work, we will exploit \textit{range-azimuth} maps obtained by 2D Capon filtering as our input features for \textit{machine learning} classification.

In section \ref{sec1}, we introduce the proposed algorithm and the system formulation based on FMCW radar and the high-resolution Capon filtering for joint range-azimuth estimation. In section \ref{sec1.1}, we briefly explain the classification. In section \ref{sec3}, we discuss on the experimental results and finally, we will conclude the paper with wrapping up the achievements and further possible extensions for the future works in section \ref{conc}.

\section{System design} \label{sec1}
\begin{figure*}[!t]
	\centering
	\includegraphics[width=7in]{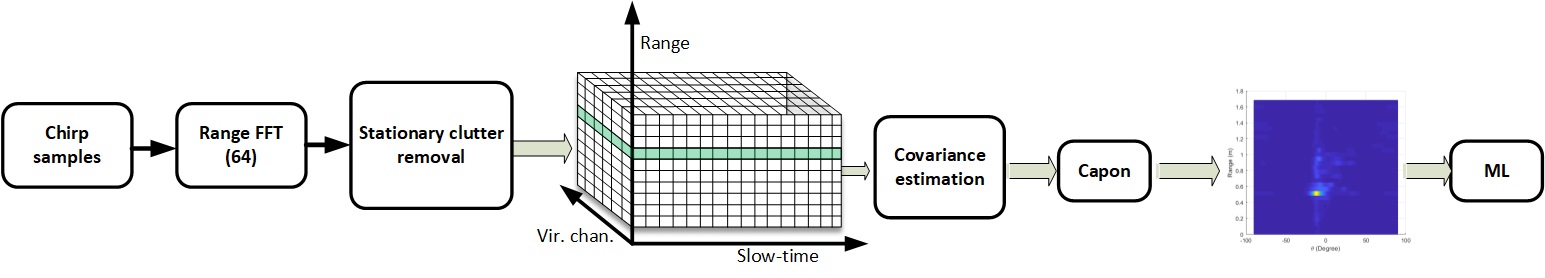}
	\caption{Vehicle occupant detection algorithm.}
	\label{vod_algorithm}
\end{figure*}
In Fig. \ref{vod_algorithm}, the proposed signal processing chain is illustrated. The \textit{range FFT} is applied on the received chirp samples. In a time division multiplexed (TDM) MIMO FMCW radar, a sequence of chirps are sent in a frame from different transmit antennas. At the receiver, the signal is collected and assigned to a virtual channel such that each channel contains the data transmitted and received from and to a unique pair of transceiver which is done in stationary clutter removal stage together with removing the average of each range bin. In fact, removing the average is equivalent to eliminating the stationary scatterers. Then, the covariance matrix of the virtual channel vector across each range bin is calculated and capon filter will be applied to estimate the angle of arrival spectrum of each range. By putting them in a matrix, \textit{range-azimuth map} will be constructed and is delivered to ML section to perform classification. 
 
\subsection{Range-azimuth map derivation} \label{sec1.1}
In a FMCW radar having up chirps i.e. only positive-slope chirps, the received baseband signal from one target at l'th antenna element is:
\begin{align}
	x_l(t_f,t_s) &= b_l \exp\Big[-j \big( 2\pi f_b t_f + 2\frac{v t_s}{\lambda_{max}} + \tau_l  \nonumber \\
	&  + \xi_l + \Delta \psi_l(t_f,t_s) \big) \Big] 
	+ e_l(t_f,t_s)
	\label{eq1}
\end{align}
where $ t_f $ and $ t_s $ are referring to two different time scales. The former refers to the time scale of a chirp period while the latter is referring to the time scale greater than a chirp period. $ b_l $, $ f_b $, $ v $, $ \lambda_{max} $, $ \tau_l $, $ \xi_l $, $ \Delta \psi_l(t_f,t_s) $, and $ e_l(t_f,t_s) $ are the channel gain, beat frequency \footnote{it contains the range of the target. }, target's velocity, the wavelength corresponding to the start frequency of the FMCW ramp, the phase shift at l'th receiver due to the angle of arrival, the channel phase, the residual phase noise existing in both $ t_f $, $ t_s $, and the additive noise, respectively. By reordering terms in \eqref{eq1} and stacking the received signal from all receiver channels in a column vector, it can be expresses as:
\begin{align}
	\boldsymbol{x} (t_f,t_s) &= \Gamma \boldsymbol{a} (\theta) y(v,t_s) s(t_f,t_s) + \boldsymbol{e} (t_f,t_s) \label{eq2}	\\
	\Gamma &  := \text{diag} \Big(b_1 \exp (-j \xi_1 ), \cdots,  b_L \exp (-j \xi_L ) \Big) \nonumber \\
	\boldsymbol{a} (\theta) & := \left[\exp \left( -j \tau_1 \right), \cdots, \exp \left( -j \tau_L \right) \right]^T \nonumber \\
	y(v,f_b,t_f,t_s) & := \exp \left(-j \left(\frac{2v t_s}{\lambda_{max}} + 2\pi f_b t_f + \Delta \psi (t_f,t_s) \right) \right) \nonumber 
\end{align} 
in which $ \theta $ is the angle of arrival (AoA) of the target. $ \Gamma $ depends on the channel gain/phase mismatches and $ \boldsymbol{a} $ depends on (AoA) and it is called the \textit{steering vector}. If there is more than one target at different ranges i.e. with different $ f_b $, then the received vector is the summation of all the vectors received from each target, thus:
\begin{align}
	\boldsymbol{x} (t_f,t_s) &= \Gamma A (\boldsymbol{\theta}) Y (\boldsymbol{v},\boldsymbol{f}_b,t_f,t_s) + \boldsymbol{e} (t_f,t_s) \label{eq3} 
%	\boldsymbol{y} & = \left[\exp \left(-j 2 \dfrac{v_1 t_s}{\lambda_{max}} \right), \cdots, \exp \left(-j 2 \dfrac{v_K t_s}{\lambda_{max}} \right) \right]^T \nonumber 	
\end{align}
where A is $ L\times K $ matrix with K is the number of targets and it has columns corresponding to the steering vector of each target. Matrix $  \boldsymbol{Y}  $ is diagonal matrix with the elements of $ y(v,f_b,t_f,t_s) $. In \eqref{eq3}, the vectors $ \boldsymbol{\theta}$, $ \boldsymbol{v} $, and $ \boldsymbol{f}_b $ are the unknown parameters of the targets; however, only elements of A are functions of receiver channel indexes. Hence, matrix $  \boldsymbol{Y}  $ does not contribute to the covariance of $ \boldsymbol{x} $. In fact, the covariance matrix of $ \boldsymbol{x} $ can be computed as the following when the additive noise is uncorrelated to  $ \boldsymbol{Y} $: 
\begin{equation}
	R = \mathbb{E} (\boldsymbol{x} \boldsymbol{x}^H) = P_s \Gamma A (\boldsymbol{\theta}) A ^H(\boldsymbol{\theta}) \Gamma^H + R_n
	\label{eq4}
\end{equation}
in which $ P_s $ is the power of $ s(t_s) $. The matrix $ R_n $ is the noise covariance and it is positive definite matrix by assuming that the noise at each receiver is independent of the others. Moreover, the first term in the covariance of \eqref{eq4}, is positive definite since $ A(\boldsymbol{\theta}) $ is a \textit{Vandermonde} matrix with positive kernels \cite{spec_prop}. Therefore, $ R $ is positive definite and it is invertible. The Capon output filter spectrum is computed as follows:
\begin{equation}
	\Phi (\boldsymbol{\hat{\theta}}) = \dfrac{1}{\boldsymbol{a}^H(\hat{\theta}) R^{-1} \boldsymbol{a}(\hat{\theta})}
	\label{eq5}
\end{equation}
where $ \hat{\theta} $ is a test unknown AoA. In an 

\subsection{Machine learning} \label{sec1.2}
\begin{table}
	\renewcommand{\arraystretch}{1.3}
	\caption{Definition of classes}
	\label{classes}
	\centering
	\begin{tabular}{|c|l|}
		\hline
		\textbf{class} & \textbf{Definition} \\
		\hline
		\hline
		\textbf{No-one} & the car is empty \\
		\hline
		\textbf{Row1}  & one person is in the first row \\
		\hline
		 \textbf{Row2}  & one person is in the second row \\
		 \hline
		 \textbf{Row3}  & one person is in the third row \\
		 \hline
		 \textbf{Row12}  & two persons, one in the first and one in the second rows \\
		 \hline
		 \textbf{Row13}  & two persons, one in the first and one in the last rows \\
		 \hline
		 \textbf{Row23}  & two persons, one in the second and one in the last rows \\
		 \hline
		 \textbf{Row123} & three persons, each on different row \\
		 \hline
	\end{tabular}
\end{table}
As Fig. \ref{vod_algorithm} shows, machine learning classification is the last stage of our proposed in-vehicle occupant detection. After finding \textit{range-azimuth} dataset, the size of it reduced by using principle component analysis (PCA) without information loss. The dataset is split to training and test datasets by the ratio of 8 to 2. The split is fair such that the size of data for each class is equal in both the training and test sets. We applied SVM, a supervised classifier, to classify the Capon filter output to 8 classes defined in Table \ref{classes}.

It is also necessary to optimize the hyperparameters utilized by the classifier to get the optimum classification. Thus, a grid search with \textit{k-fold cross validation} is employed to search for the best hyperparameters that will increase the prediction accuracy of the classifiers. For evaluation, we tested the model using 5-fold cross validation. 

\section{Experimental results} \label{sec3}

\begin{table}
	\renewcommand{\arraystretch}{1.3}
	\caption{FMCW Radar configuration}
	\label{radar_param}
	\centering
\begin{tabular}{|c|c|c|c|c|c|}
	\hline
	 Parameter & K$ ^\mathtt{a} $ & $ T_r ^\mathtt{b}$ & $ B_s ^\mathtt{c}$ & $ f_r ^\mathtt{d}$ & $ f_{b, max} ^\mathtt{e}$\\
	\hline
	\hline
	Value & $ 98 MHz/\mu s $ & $ 580 \mu s $ & 3920 MHz & 6.25/s & 2.2 MHz \\
	\hline
	\multicolumn{2}{c} {$ ^\mathtt{a} $ chirp slope} & \multicolumn{2}{c}{$ ^\mathtt{b} $ chirp duration}  & \multicolumn{2}{c}{$ ^\mathtt{c} $ sweeping BW} \\
	\multicolumn{2}{c}{$ ^\mathtt{d} $ frame rate} & \multicolumn{2}{c}{$ ^\mathtt{e} $ ADC sampling rate}
\end{tabular}
\end{table}
\begin{figure}[t]
	\centering
	\includegraphics[width=3in]{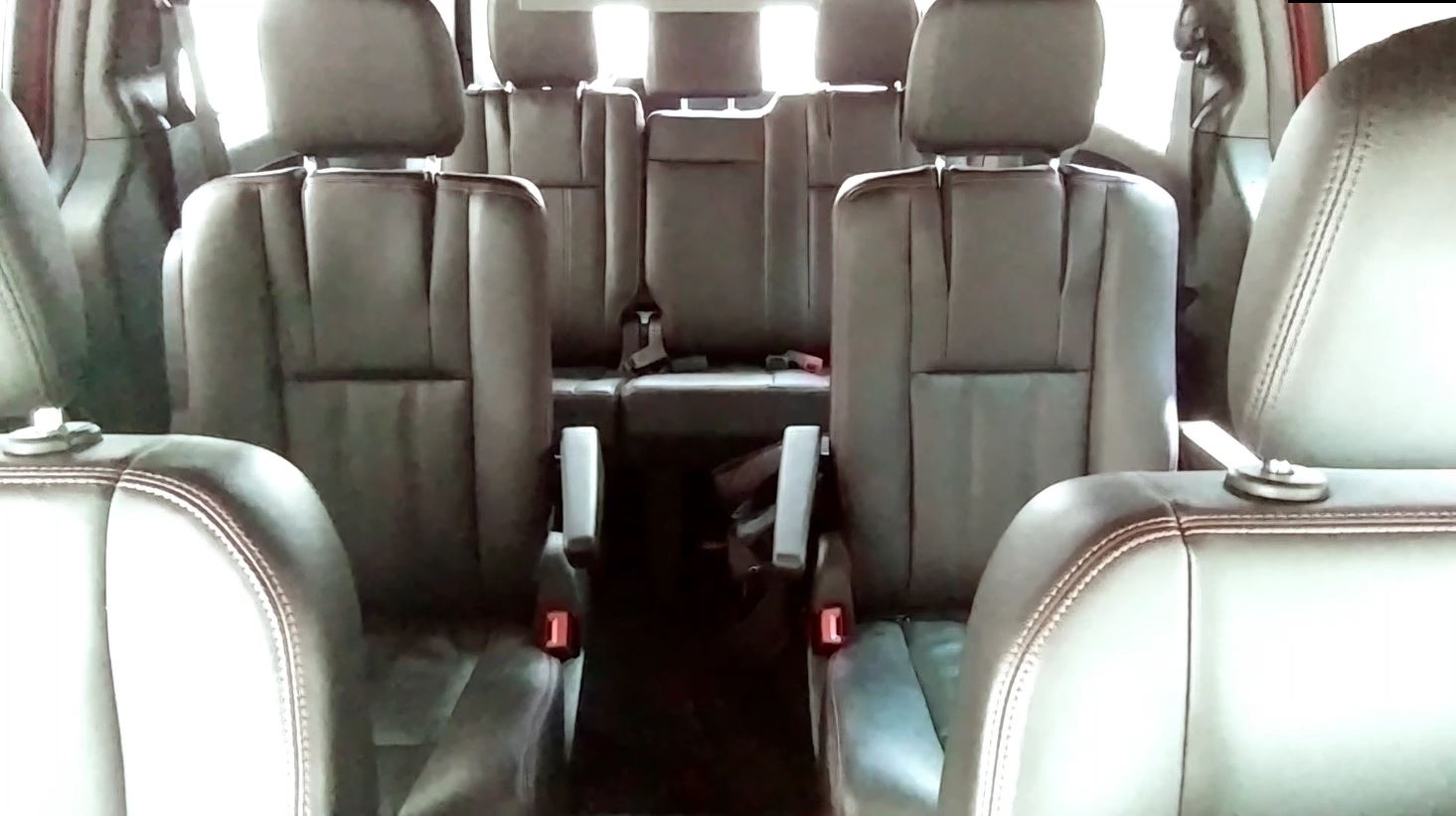}
	\caption{Minivan indoor look}
	\label{minivan}
\end{figure}
\begin{figure}
	\centering
	\includegraphics[width=3in]{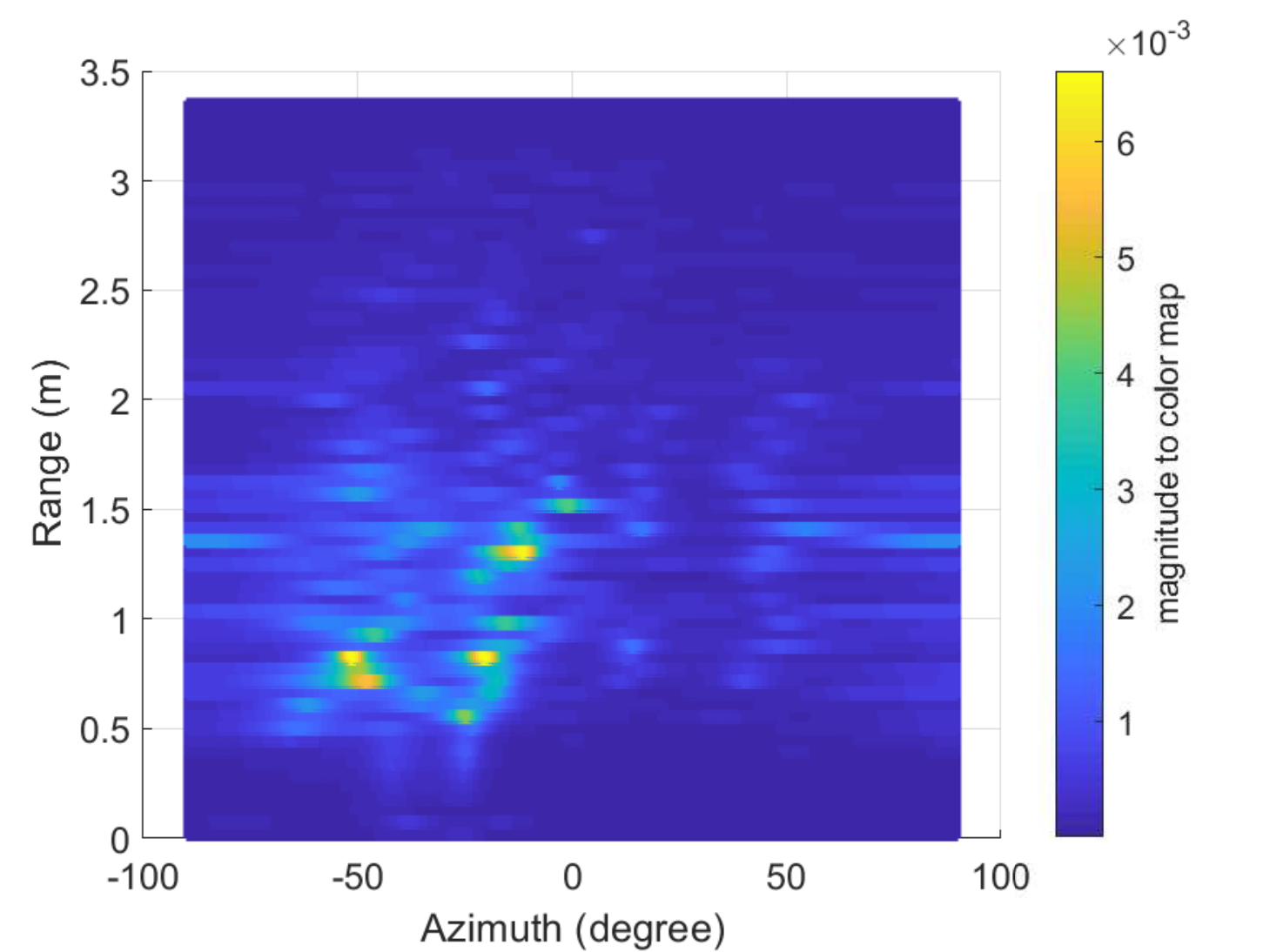}
	\caption{Range-azimuth map when the seat number 1, 4, and 5 are occupied. }
	\label{rng_az_map}
\end{figure}

We used \textit{Texas Instrument} (TI) mm-wave FMCW chip for our experiments which operates at 77GHz. In addition, there were three rows in \textit{minivan} as shown in Fig. \ref{minivan}. The seats are numbered starting from the driver and each row from right to left of the figure. During recordings, passengers were allowed to freely move their hands, talk to each other, and work with their phones. We consider eight classes, empty car and either a row is occupied or not (Table \ref{classes}). In fact, our class definition is a small collection of all possible situations in a car;  otherwise, the total number of seat occupations is $ 2^7=128 $. The recording duration is the same for all classes in order to have unbiased training.

TI radar chip has 3Tx and 4Rx and two transmitters were used in TDM MIMO mode to construct eight virtual receivers (further radar parameters are listed in Table \ref{radar_param}). Thus, for azimuth detection, we have 8 spatial samples resulting in $ 2/8 = 0.25 \; Radian$ or 14-degree resolution. This means that at 1 meter away from the radar, two targets can be resolved only if they are separated more than 25 cm; otherwise, the targets are spatially correlated. For instance, in Fig. \ref{rng_az_map}, represents a sample \textit{range-azimuth} map obtained from \eqref{eq5} when the driver (seat number 1) and two other passengers are in the minivan (see Fig. \ref{minivan}). The figure does not show clear visual passenger separation due to low angular resolution. However, SVM classifier can identify this map with three occupied rows.  

\begin{figure}[t]
	\centering
	\includegraphics[width=3.5in]{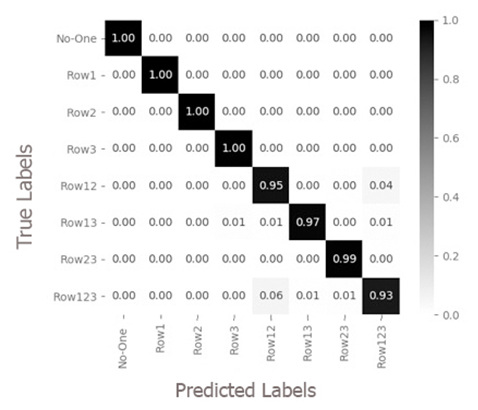}
	\caption{Confusion matrix of SVM}
	\label{conf_mat}
\end{figure}

Confusion matrix of SVM classifier is shown in Fig. \ref{conf_mat}. From Fig. \ref{conf_mat}, when the car is vacant, there is no false alarm. Additionally, if the car is not empty, SVM correctly indicates that there is at least one person (see the first column of Fig. \ref{conf_mat}). Moreover, in the presence of an occupant; i.e. \textit{row1, row2, row3}; SVM certainly identifies the case. However, most of the misclassifications occurred for localization of more than one occupant in the car i.e. \textit{row123} case. Finally, by using 5-fold cross validation, 97.8\% correct detection rate is obtained.
	
\section{Conclusion} \label{conc}
In this paper, we investigated the in-vehicle occupancy detection with a mm-wave FMCW radar. Also, we addressed the low angular resolution which limits the visual perception of the occupant location. However, with the aim of machine learning classifier, namely SVM, a high accuracy was obtained to determine one of the defined scenarios. Therefore, although the resolution is low, by defining efficient features the accuracy can be enhanced. 

\section*{Acknowledgment}
The authors acknowledge financial support of Natural Sciences and Engineering Research Council of Canada (NSERC) and Ontario Centres of Excellence (OCE).



\bibliographystyle{IEEEtran}
\bibliography{ConfRefs}
%\begin{thebibliography}{1}


%\end{thebibliography}

\end{document}